\documentclass{jfm}
\pdfoutput=1 
\usepackage{graphicx}
\usepackage{epstopdf, epsfig}
\let\proof\relax

\usepackage{amsmath,amsfonts,amssymb,amsthm,color}
\usepackage[round]{natbib}
\usepackage[utf8]{inputenc}
\usepackage[normalem]{ulem}
\usepackage{soul}
\usepackage[font=normal]{subcaption}
\usepackage{graphicx}

\shorttitle{A buoyancy-driven translating pancake droplet in a Hele-Shaw cell}
\shortauthor{I. Shukla,  N. Kofman, G. Balestra, L. Zhu and F. Gallaire}

\title{Film thickness distribution in gravity-driven pancake-shaped droplets rising in a Hele-Shaw cell}
\author{Isha Shukla\aff{1}, Nicolas Kofman\aff{1}, Gioele Balestra\aff{1}, Lailai Zhu\aff{1,2,3} \and Fran\c cois Gallaire\aff{1} \corresp{\email{francois.gallaire@epfl.ch}}}

\affiliation
{\aff{1}Laboratory of Fluid Mechanics and Instabilities, \'{E}cole Polytechnique F\'{e}d\'{e}rale de Lausanne, Lausanne, CH-1015, Switzerland
\aff{2}Department of Mechanical and Aerospace Engineering, Princeton University, Princeton, New Jersey 08544, USA
\aff{3}Linn\'{e} Flow Centre and Swedish e-Science Research Centre (SeRC), KTH Mechanics, Stockholm SE-10044, Sweden}

\begin{document}
\maketitle
\begin{abstract}
We study here experimentally, numerically and using a lubrication approach; the shape, velocity and lubrication film thickness distribution of a droplet rising in a vertical Hele-Shaw cell. The droplet is surrounded by a stationary immiscible fluid and moves purely due to buoyancy. A low density difference between the two mediums helps to operate in a regime with capillary number $Ca$ lying between $0.03-0.35$, where $Ca=\mu_o U_d /\gamma$ is built with the surrounding oil viscosity $\mu_o$, the droplet velocity $U_d$ and surface tension $\gamma$. The experimental data shows that in this regime the droplet velocity is not influenced by the thickness of the thin lubricating film and the dynamic meniscus. For iso-viscous cases, experimental and three-dimensional numerical results of the film thickness distribution agree well with each other. The mean film thickness is well captured by the \cite{aussillous2000quick} model with fitting parameters. The droplet also exhibits the ``catamaran'' shape that has been identified experimentally for a pressure-driven counterpart~\citep{huerre2015droplets}. This pattern has been rationalized using a two-dimensional lubrication equation. In particular, we show that this peculiar film thickness distribution is intrinsically related to the anisotropy of the fluxes induced by the droplet's motion.  
\end{abstract}
\section{Introduction} \label{introduction}
Transport of droplets and bubbles in confined environments is a common process in engineering applications, such as microscale heat transfer and cooling using a slug flow~\citep{kandlikar2012history, magnini2013numerical}, enhanced oil recovery based on foam injections where bubbles move in porous media~\citep{farajzadeh2009investigation} and microfluidic engineering using, droplets as micro-reactors~\citep{song2006reactions} and cell-encapsulating micro-compartments~\citep{he2005selective}, to name a few. The study of transported droplets in confined dimensions also extends to biological science where red blood cells traversing passages with non-axisymmetric geometries were analysed \citep{halpern1992squeezing}.

Pioneering work has been initiated for a long bubble translating inside a straight cylindrical tube by \cite{taylor1961deposition} conducting experiments and \cite{bretherton1961motion} combining experiments and asymptotic analysis. The analysis of Bretherton showed that the lubrication equations, at a very small capillary number $Ca$, were similar to their one-dimensional version assuming spanwise invariance. He established the famous asymptotic relation between the uniform film thickness $H_{\infty}$ and the capillary number in the $Ca < 10^{-3}$ regime, namely, $H_{\infty}/W = P(3Ca)^{2/3}/2$, where $W$ is the tube diameter and $P$ a coefficient. The capillary number $Ca = \mu_o U_d/\gamma$ is built with the carrier phase dynamic viscosity $\mu_o$, the droplet velocity $U_d$ and the surface tension $\gamma$ between the two fluids. 
\cite{aussillous2000quick} proposed 
\begin{equation}
\frac{H_{\infty}}{W} = \frac{1}{2}\frac{P(3Ca)^{2/3}}{1+PQ(3Ca)^{2/3}},
\end{equation} 
as the \textit{Taylor's law} including an empirical coefficient $Q=2.5$, with the coefficient $P$ inherited from \cite{bretherton1961motion}; this law was validated against the experimental data of \cite{taylor1961deposition} for $Ca < 2$. The empirical relation was rationalised by incorporating into the analysis of Bretherton the so-called ``tube-fitting''condition, namely, that the bubble-film combination should fit inside the tube~\citep{klaseboer2014extended}. Besides those work considering the steady translation, \cite{yu2018time} has recently investigated how the lubrication film evolves between two steady states of a Bretherton bubble by combining theory, experiments and simulations.

Contrary to the translating bubble in a capillary tube, a bubble moving in a Hele-Shaw cell (two closely gapped parallel plates) resembles a flattened pancake. This configuration is relevant to microfluidic applications \citep{baroud2010dynamics} where the thickness of the microfluidic chips is much smaller than their horizontal dimension. Owing to the mathematical similarity between the governing equations of the depth-averaged Hele-Shaw flow and those of the two-dimensional (2D) irrotational flow as proved by \citet{stokes1898mathematical} and commented by \citet{lamb1993hydrodynamics}, potential flow theory was adopted to study the motion of a Hele-Shaw bubble theoretically \citep{taylor1959note} and numerically \citep{tanveer1986effect}. Based on the stress jump derived by \citet{bretherton1961motion} and \citet{park1984two}, 2D depth-averaged simulations including the leading-order effects of the dynamic meniscus were also carried out \citep{meiburg1989bubbles}.

Motivated by the applications of droplet-based microfluidics, several works have been recently conducted to investigate the dynamics of a pressure-driven Hele-Shaw droplet. \cite{huerre2015droplets} and \cite{reichert2018topography} performed high-precision experiments using reflection interference contrast microscopy technique to study the pressure-driven droplets, observing the so-called ``catamaran'' droplet shape. Simulations based on a finite volume method \citep{ling2016droplet} and a boundary integral method (BIM) \citep{zhu2016pancake} were carried out, confirming such a peculiar interfacial feature. It has to be mentioned that the much earlier work of \citet{burgess1990analysis} performing a multi-region asymptotic analysis subtly revealed this feature for a Hele-Shaw bubble, which was rather unnoticed.

Limited work has been conducted for the gravity-driven droplets in a Hele-Shaw cell. \cite{eri2011viscous} and \cite{yahashi2016scaling} studied experimentally such configurations, trying to build up the scaling laws for the viscous drag friction of the Hele-Shaw droplets. Recently, \cite{keiser2018dynamics} conducted experiments to study a sedimenting Hele-Shaw droplet, focusing on its velocity as a function of confinement, viscosity contrast and the lubrication capacity of the carrier phase. 

In this work, we combine experiments, simulations and a lubrication model solved numerically to study the  buoyancy-driven translation of a droplet inside a vertical Hele-Shaw cell. We examine the droplet velocity, film thickness and how they vary with the density and viscosity difference between the droplet and carrier phase. We introduce the experimental setup in \S \ref{expSetup}, followed by the experimental results of the droplet mean velocity and film thickness in \S\ref{expResults} and \S\ref{expResultsH} respectively. The comparison between the three-dimensional (3D) BIM simulations and the experiments is shown in \S\ref{NumericsComp}. The lubrication equation employed to model the problem is presented in \S\ref{pancakes_tx:2D} where the numerical solution of the lubrication equation is compared to the 3D simulation results in \S \ref{2dnonlinear}. The film thickness pattern is rationalised by solving the linearised 2D lubrication equation, which is presented in \S\ref{pancakes_tx:2Dlin}. We finally summarise our results in \S\ref{conclusion}  with some discussions.
\section{Experimental setup} \label{expSetup}
\begin{figure}
\centering
\includegraphics[trim=60 0 0 20,clip,width=1\linewidth]{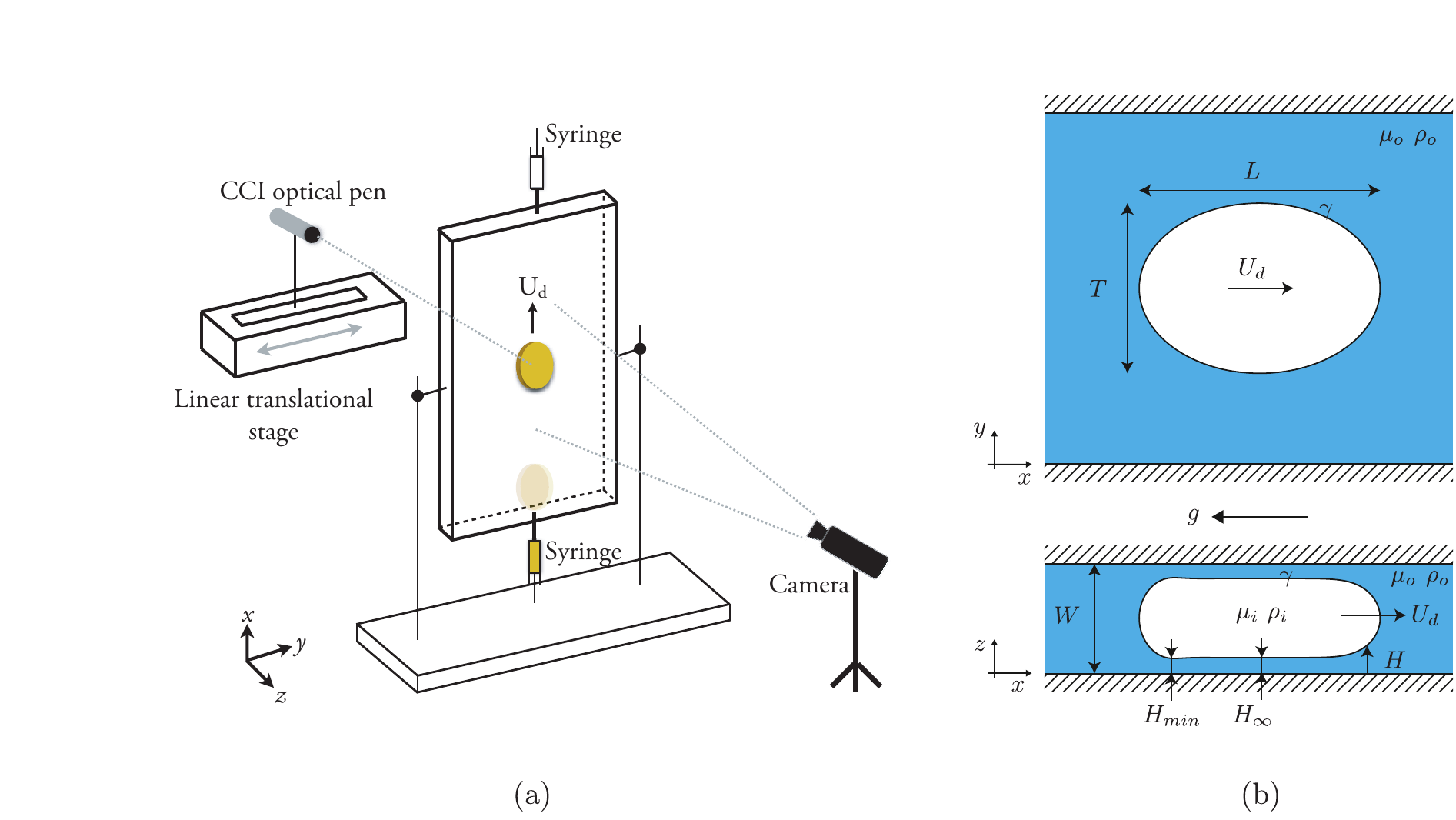}
\caption{(a) Schematic of the experimental set-up. (b) Sketch of the 
problem: a droplet with density $\rho_i$ and dynamic viscosity $\mu_i$ moving at velocity $U_d$ in a 
Hele-Shaw cell of height $W$, where the carrier phase has a dynamic viscosity $\mu_o$ and its density $\rho_o 
>\rho_i$. The in-plane ($x, y$) projection shows the drop's longitudinal and transversal lengths, $L$ and $T$, respectively. The out-of-plane ($x, z$) drop shape shows the thickness $H_{\infty}$ of the uniform thin film and  
the minimum thickness $H_{min}$ of the film along the centreline. }
\label{fig:schematic}
\end{figure}
A vertical Hele-Shaw cell made of two parallel glass plates, separated by a gap $W$, is filled with silicone oil of dynamic viscosity 560 mPa\ s and density 972 kg\ m$^{-3}$, measured at $20^{\circ}$C. An oil drop is injected into the silicone oil medium from the bottom using a syringe as shown in figure \ref{fig:schematic}(\textit{a}). The drop moves as a result of buoyancy. The higher the density difference between the inner and outer medium, the higher the drop velocity $U_d$. The spanwise and streamwise cell dimensions are sufficiently large compared to the drop size to avoid any finite size effects from the lateral walls. On the other hand, the droplet is highly confined in the wall normal direction. The droplet radius $a$ is always larger than the cell gap $W$. Given the compliance of the glass walls, the thickness of which is bounded by our optical measurement tools, the cell gap $W$ lies in the range of $[4.59-4.8]$\ mm and is recorded every time before the drop injection (see table \ref{table1}).

The injected oils are tested beforehand to ensure non-wetting conditions for the oil droplet on the cell plates. The outer silicone oil totally wets the glass plate and forms a thin film of thickness $H$, between the drop and the glass's interface (see figure \ref{fig:schematic}(\textit{b})), which is measured using a CCI optical pen (see details in Appendix \ref{CCI}). The pen is either placed fixed such that it measures the film thickness only along the centreline $L$ (\textit{centreline film thickness}) or is mounted on a linear translational stage to perform lateral scans while the drop moves longitudinally. With an acquisition frequency of 200-500 Hz, scanning amplitude of 20-30 mm and frequency of 2-3 Hz, the obtained experimental data are interpolated in Matlab to obtain the film thickness maps for the entire drop. Droplet size and velocity determine the optimal acquisition frequency for the thickness sensor, and the scanning amplitude and frequency for the linear translational stage. 

We observe that for the chosen inner oil volumetric range, the droplet in-plane shape is no longer a circle but closer to an oval, hence we refer to the drop longitudinal length (along the direction of gravity) as $L$ and to the transverse length as $T$, as shown in figure \ref{fig:schematic}(\textit{b}). The drop motion is captured using a Phantom Miro M310 camera with a Nikon 105 mm macro lens. The spatio-temporal analysis of the movie ensures uniform drop velocity as the drop moves along a longitudinal distance of $5L$ or more. 

The drop volume $Q$ is expressed as a pancake of radius $a$ and height {$W$-$2H_{\infty}$, where $H_{\infty}$ is the mean film thickness. We can simplify $Q$ as $\pi a^2W$ when $H_{\infty} \ll W$. For the volumetric range used for the inner oils, we found that the longitudinal and transverse lengths, $L$ and $T$, scale as the pancake radius $a$. The aspect ratio $\alpha$ is expressed as the ratio $a/W$. Keeping the cell gap $W$ fixed, data for different aspect ratios are obtained using three different volumes (0.5ml, 1ml, 1.5ml) for each oil. 

Six oils with physical properties as mentioned in table \ref{table2} are used. The surface tension $\gamma$ between the inner and outer medium is measured using Teclis tensiometer and the oil viscosity and density are measured using Anton Paar SVM$^{\text{TM}}$ 3000 viscometer. The experiment is performed at $20^{\circ}$C-$22^{\circ}$C.
\begin{table}
  \begin{center}
    \def~{\hphantom{0}}
      \begin{tabular}{llll}
Inner oil & $\mu_i$ (mPa\ s) & $\rho_i$ (kg\ m$^{-3}$) & $\gamma$ (mN\ m$^{-1}$) \\ [0.5ex] 
Linseed oil& 49 & 929 & 2.88 \\ 
Sunflower oil& 69.5 & 921.5 & 2.73\\
Sesame oil& 71 &919.1  & 2.64 \\ 
Olive oil& 79.3 & 913.3& 2.55 \\ 
Peanut oil& 83.8 & 913.3 & 3.11  \\ 
Ricin oil (type 1) + $10\%$ ethanol& 302 & 943.1& 4.97 \\ 
Ricin oil (type 2) + $10\%$ ethanol& 322 & 943.3& 4.46 \\ 
       \end{tabular}
\caption{Dynamic viscosity $\mu_i$ and density $\rho_i$ of the inner oils.
Those of the outer oil are $\mu_o=560$\ mPa\ s and $\rho_o=972$\ kg\ m$^{-3}$ except for the case of Ricin (type 2) + $10\%$ ethanol for which the outer oil has $\mu_o=319$ mPa\ s, $\rho_o=970.5$\ kg\ m$^{-3}$. The interfacial surface tension between the inner-outer oils is $\gamma$. These properties were obtained at a room temperature 20$^\circ$C.}
    \label{table2}
  \end{center}
\end{table}

The ratio $\lambda$ between the dynamic viscosity of the inner and outer phase lies between $[0.09-0.54]$. 
In addition to this range, another set of experiments is performed with $\lambda = 1.01$, where the outer medium is silicone oil ($\mu_o=319$ mPa\ s, $\rho_o=970.5$\ kg\ m$^{-3}$) and the inner medium is a mixture of ricin oil and $10\%$ ethanol ($\mu_i=322$ mPa\ s, $\rho_i=943.3$ kg\ m$^{-3}$) for three different drop volumes. The interfacial surface tension between these oils is 4.46\ mN\ m$^{-1}$. Notations for physical parameters and their definitions are detailed in table \ref{table1}. 
\begin{table}
\begin{center}
\def~{\hphantom{0}}
\begin{tabular}{llll}
Symbol & Definition & Expression & Working range \\ [0.5ex] 
 $W$ & cell gap & - & 4.59-4.8 mm \\ 
$U_d$ & drop velocity & - & 0.4-1.6 mm\ s$^{-1}$ \\
$Q$ & injected drop volume & - & 0.44-1.5 ml \\ 
$a$ & pancake equivalent radius & $\sqrt{Q/\pi W}$ & 5.8-10.3 mm \\ 
$\alpha$ & aspect ratio & $a/W$ & 1.25-2.27 \\ 
$\Delta \rho$ & density difference & $|\rho_i-\rho_o|$ & 27.3-58.8 kg\ m$^{-3}$ \\ 
$\mu_i$ & dynamic viscosity - droplet &  - & 49-322 mPa\ s \\ 
$\mu_o$ & dynamic viscosity - outer medium & - & 319-560 mPa\ s \\ 
$\gamma$ & interfacial surface tension & - & 2.5-5.0 mN\ m$^{-1}$ \\ 
$\lambda$ & dynamic viscosity ratio & $\mu_i/\mu_o$ & 0.09-1.01 \\ 
$Ca$ & capillary number & $\mu_o U_d/\gamma$ & 0.03-0.35  \\
$Bo$ & Bond number & $\Delta \rho g a^2/\gamma$ & 1.8-23.3 \\
\end{tabular}
\caption{List of notation, definition and working range.}
\label{table1}
\end{center}
\end{table}
\section{Experimental acquisition of the drop characteristics and their comparison with 3D BIM simulations}\label{ExpNumRes}
\subsection{Experimental results for drop velocity}\label{expResults}
Considering the bulk dissipation only, the resulting viscous drag force acting on the drop scales as $F_d \sim (\mu_i +\mu_o) U_d \pi a^2 {W}^{-1}$.  Unlike \citet{okumura2017viscous}, we consider both the inner and outer viscosities since they are of the same order. Balancing the total drag force with the buoyancy force, $F_g \sim \Delta \rho g \pi a^2 W$, we obtain a scaling for the droplet mean velocity as
\begin{equation} \label{eq:UdScaling}
U_d \sim \frac{\Delta \rho g W^2}{ (\mu_i +\mu_o)} ,
\end{equation}
where $\Delta \rho$ is the density difference and $g=9.81$ m\ s$^{-2}$. 

Under the assumption of cylindrical penny-shaped wetting drops, a theoretical expression for the drop velocity can be obtained from \citet{maxworthy1986bubble}, \citet{bush1997anomalous} and \citet{gallaire2014marangoni}. \citet{gallaire2014marangoni} deduced the drop velocity in a Hele-Shaw cell, subjected to both buoyancy and Marangoni flow, using depth-averaged Stokes equations, called the Brinkman equations. In the absence of Marangoni effect and at leading order, \citet{bush1997anomalous} and \citet{gallaire2014marangoni} predicted the mean drop velocity as
\begin{equation} \label{eq:Udexact}
U_d =\frac{\Delta \rho g W^2} {12 \mu_o(\lambda+1)} .
\end{equation}
Introducing the Bond number $Bo$ (refer table \ref{table1}) we can rewrite equation \eqref{eq:Udexact} using the aspect ratio $\alpha$ as,
\begin{equation}
12\alpha^2(\lambda+1)Ca = Bo .
\label{eq:CaBoAspect}
\end{equation}

\begin{figure}
 \centering
\includegraphics[width=0.8\linewidth]{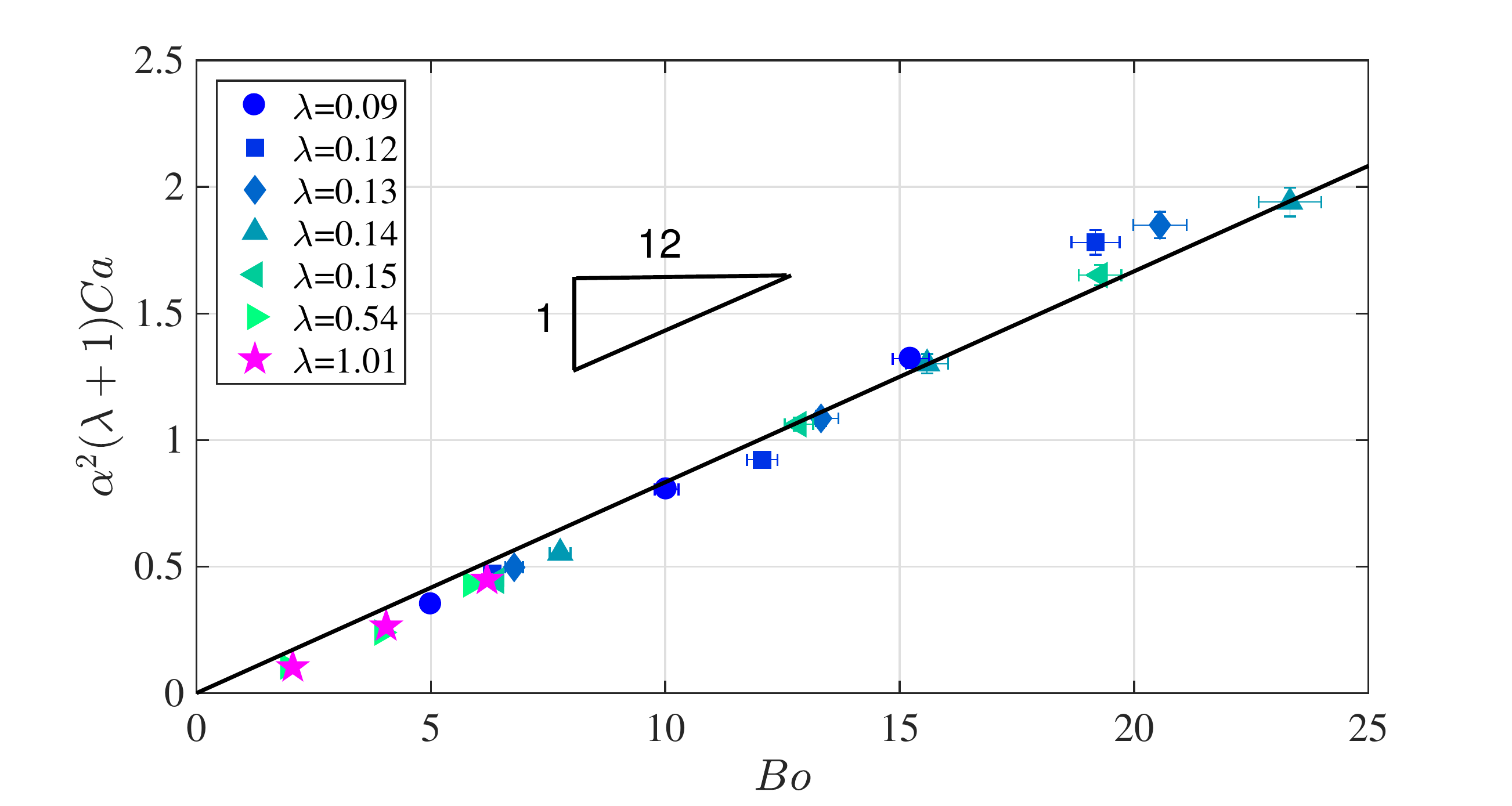}
\caption{Experimental data $\alpha^2(\lambda+1)Ca$ versus the Bond number $Bo$, where  $Ca\in [0.03, 0.35]$. The markers correspond to different viscosity ratios $\lambda$ of the inner-outer medium. The data closely fits equation 
\eqref{eq:CaBoAspect} represented by the straight line.}
\label{fig:exp1}
\end{figure} 

The experimental data {are} plotted against the theoretical equation \eqref{eq:CaBoAspect} in figure \ref{fig:exp1}. Following the trend predicted by equation \eqref{eq:CaBoAspect}, figure \ref{fig:exp1} signifies the dominant forces in play are buoyancy and viscous drag due to the volume of fluid displaced by the drop. The dissipation induced in the thin film as well as the one in the dynamic meniscus region are found not to play a role in the selected parameter range. However it has been observed that for low $Ca$ ranges, the dissipation in the thin film \citep{keiser2018dynamics} and in the dynamic meniscus \citep{reyssat2014drops} have to be taken into account. 
\subsection{Experimental results for film thickness}\label{expResultsH}
Film thickness maps were measured for different droplet velocities. Since the thickness sensor fails to capture the data in presence of high thickness gradient, no data is acquired along the drop edges, as shown in figure \ref{fig:expMap}(\textit{d}), where the black curve represents the drop in-plane boundary. For different aspect ratios, qualitatively similar thickness maps were obtained, with a high film thickness on the front edge, a constant film thickness in the centre and very low film thickness along the lateral edges of the drop, overall resembling a catamaran-like shape. The spanwise and streamwise cut made along the film thickness are shown in figure \ref{fig:expMap}(\textit{b}-\textit{c}). The centreline film thickness indicated by the streamwise cut at $y=0$ (figure \ref{fig:expMap}(\textit{b})) clearly shows a monotonically decreasing film thickness pattern, followed by a region of constant film thickness $H_{\infty}$ which then reaches a minimum value of $H_{min}$. At the rear of the droplet, the strong thickness gradient reverses the direction to have an increasing thickness profile close to the drop receding edge, thus posing technical issues to capture the film thickness. 
\begin{figure}
\centering
\includegraphics[trim=0 30 0 0,clip,width=0.8\linewidth]{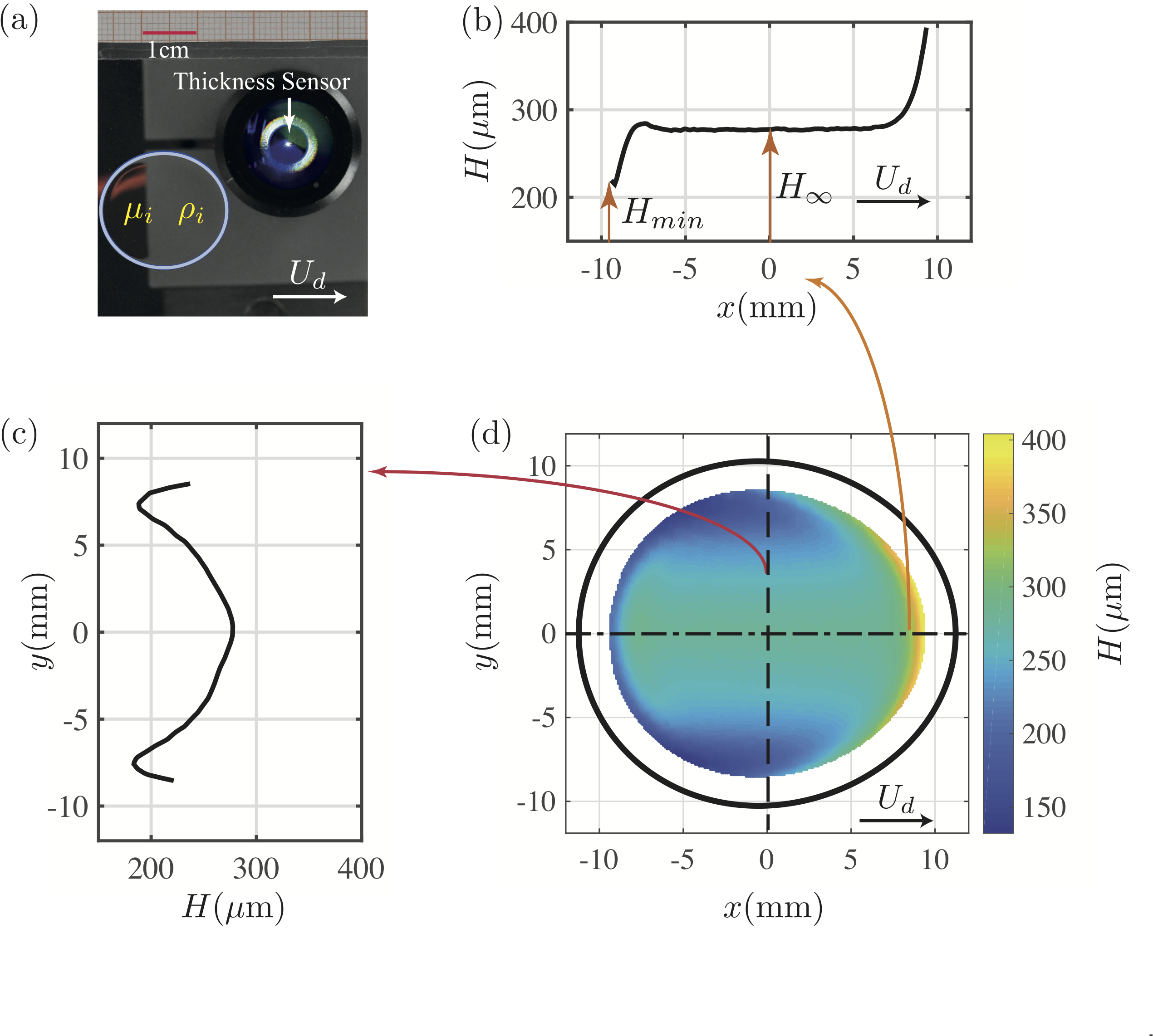}
\caption{Drop characteristics for droplet with $\lambda \sim 1$ moving with mean velocity $U_d=0.64$mm/s, $Ca=4\times10^{-2}$ and $Bo=6.2$.  (a) The blue curve shows the in-plane drop shape fitting based on equation \eqref{exp:shapePancake} with $L/2=11.22$ mm, $T/2=10.21$ mm and fitting coefficient $c=-7.485\times10^{-6}$ mm$^{-1}$. The film thickness in the streamwise and spanwise directions $y=0$ and $x=0$ of the drop are shown in (b) and (c) respectively. Figure (b) shows the typical centreline thickness profile with monotonic decreasing thickness, followed by constant thickness $H_{\infty}$ and ending with the minimum film thickness $H_{min}$. (c) The film thickness profile along the spanwise direction highlights the two minima along the lateral edge of the drop at $y\sim \pm 7.5$ mm which are clearly noticed in (d) where we see the in-plane shape in black and the obtained film thickness map. The data is missing along the drop boundaries due to the presence of high thickness gradient that cannot be captured by the thickness sensor. }
\label{fig:expMap}
\end{figure}%

A similar centreline film thickness profile was obtained for all the droplets with a distinct value of $H_{\infty}$ and $H_{min}$. These profiles are very similar to the ones of \cite{bretherton1961motion} for pressure-driven droplets, and as already noted in other works of pancakes (\citealp{huerre2015droplets}, \citealp{zhu2016pancake} and \citealp{reichert2018topography}). Nondimensionalising the mean and minimum values along the centreline using the cell gap $W$ and plotting them as a function of $Ca$ shows a saturating trend for higher $Ca$ (figure \ref{fig:exp3}). The experimental data are fitted based on the \textit{Taylor's law} model \citep{taylor1961deposition, aussillous2000quick}, according to which apart from the static and dynamic meniscus regions, the lubrication film has a constant thickness of $H_\infty$ given as:
\begin{equation} \label{HAsympExp}
\frac{H_\infty}{W}=\frac{1}{2}\frac{P(3Ca)^{2/3}}{1+PQ(3Ca)^{2/3}},
\end{equation}
where the coefficients $P=0.544$ and $Q=2.061$ are obtained from the best fit curve for the experimental data. The nonlinear equation \eqref{HAsympExp} is fitted using the Matlab function \textit{sseval} such that the objective function, defined as the sum of squared errors between the real data of $H_\infty/ W$ and the one predicted by equation \eqref{HAsympExp}, using any pair of parameters $P$ and $Q$, is the minimum. The L2 error norm for $H_\infty /W$ between the fitted and actual data is $0.02$. 

The fitting coefficients compare well with \citet{aussillous2000quick} and \citet{klaseboer2014extended}, where the mean film thickness model for bubbles ($\lambda=0$) is based on the \textit{Taylor's law} with coefficient $P=0.643$, and $Q=2.5$ and $2.79$ respectively. Fitting coefficients obtained from a 3D BIM simulation of \citet{zhu2016pancake} for pressure-driven flows and $\lambda=1$ show the same order of magnitude as the experimental ones, with $P=0.6$ and $Q=1.5$. 

In figure \ref{fig:exp3}(\textit{a}), we see that our experimental data for mean film thickness are bounded by the predicted values for the two extreme viscosity ratios of $\lambda=0$ and $\lambda=1$. Comparing the thickness predictions by \cite{klaseboer2014extended} and \cite{zhu2016pancake} for $Ca=0.1$ we see that the thickness variation is $20\%$ as $\lambda$ increases from 0 to 1. This is consistent with the $18\%$ (approximately) increase as reported in \cite{martinez1990axisymmetric} for pressure-driven drops in an axisymmetric tube. Further, this variation in thickness  reduces to a merely  $11\%$  for $Ca=0.05$, as $\lambda$ changes from 0 to 1.

The same model when used for fitting the minimum film thickness profile $H_{min}/W$ gives fitting coefficients $P=0.372, Q=1.247$ with an L2 error norm between the fitted and actual data as $0.025$. 
\begin{figure}
	\centering
	\begin{subfigure}{0.49\textwidth} 
		\includegraphics[trim=0 0 50 0,clip,width=\textwidth]{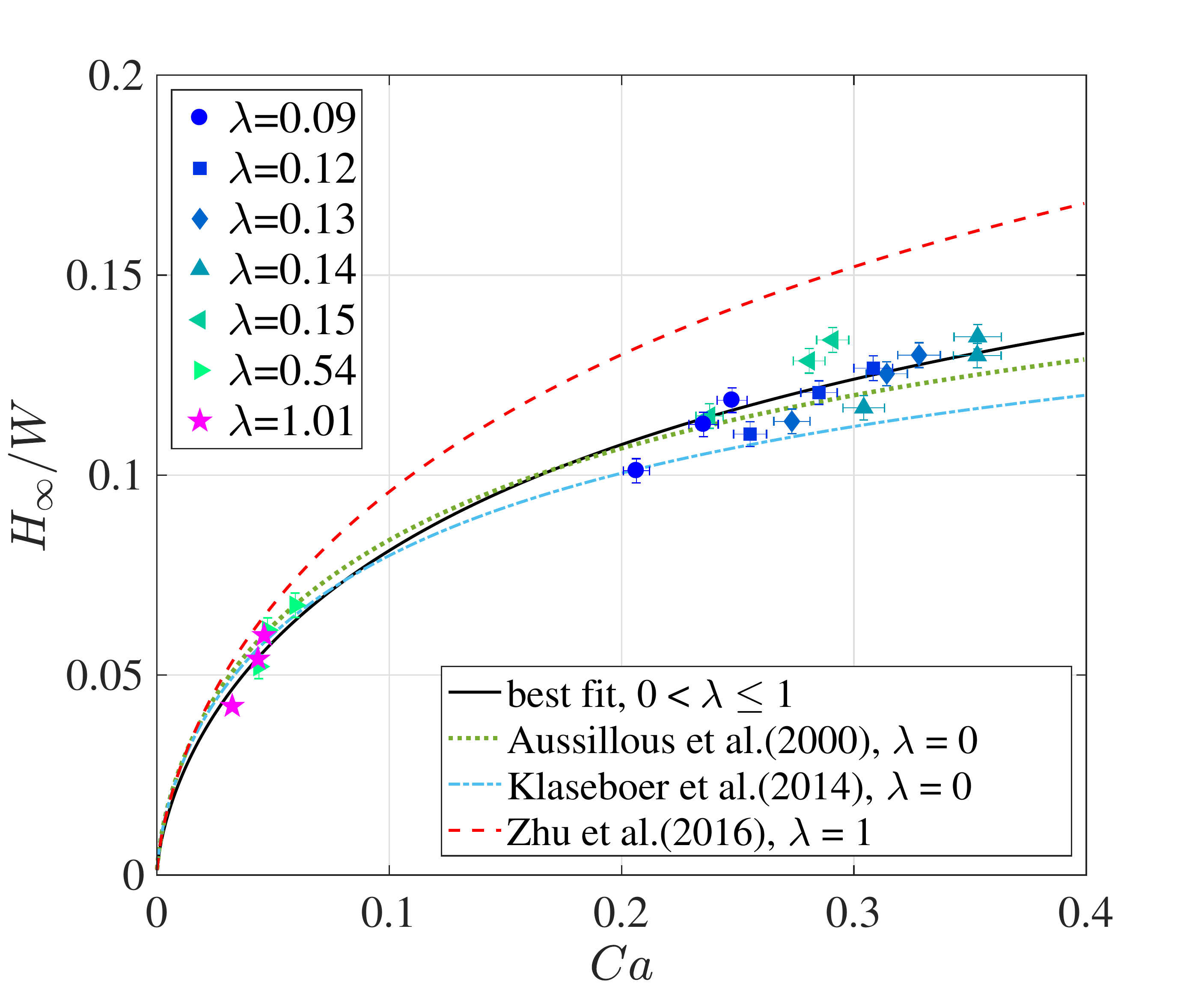}
		\caption{} \label{fig:hmean}
	\end{subfigure}
	\begin{subfigure}{0.49\textwidth} 
		\includegraphics[trim=0 0 50 0,clip,width=\textwidth]{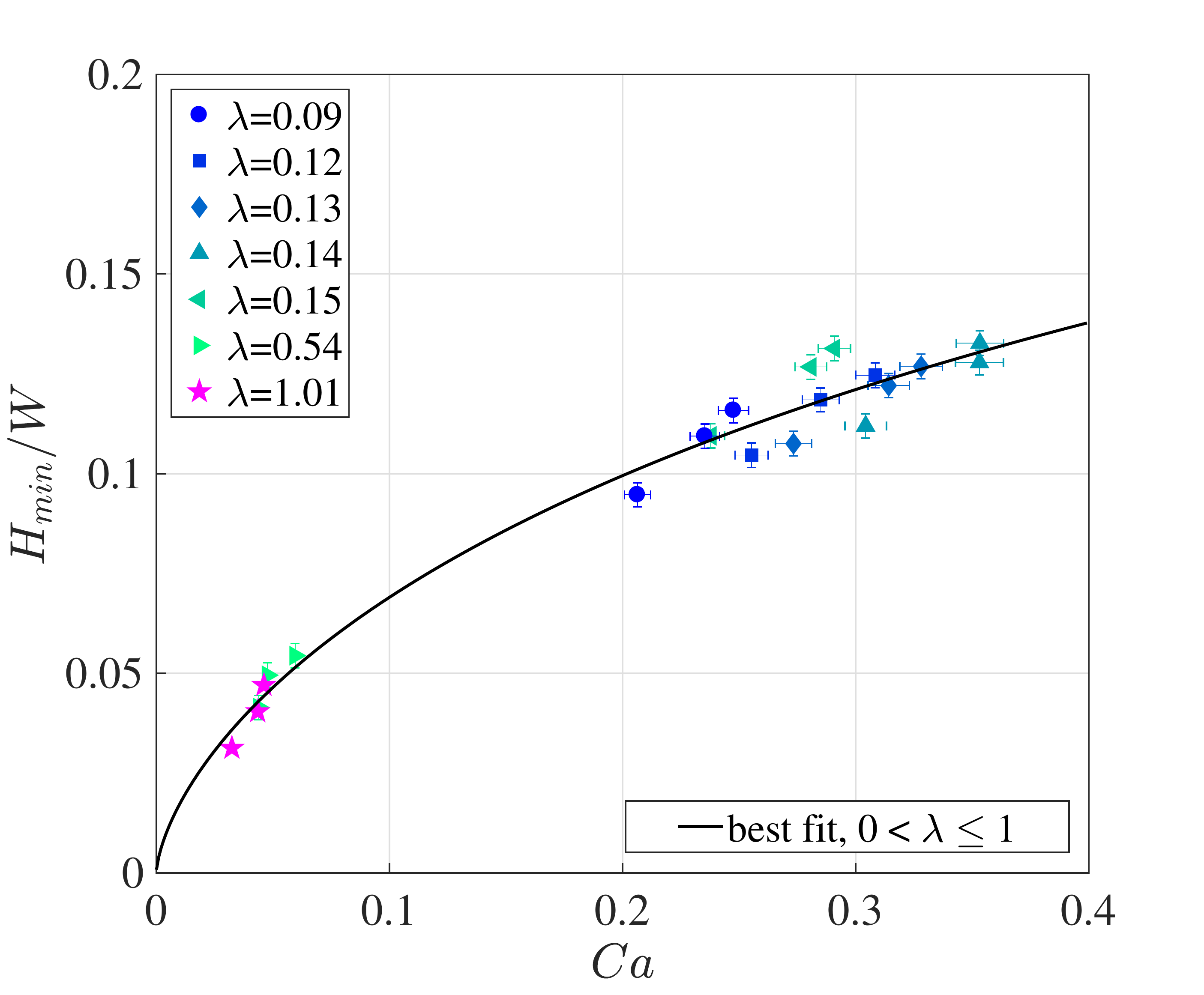}
		\caption{} \label{fig:hmin} 
			\end{subfigure}
		\caption{(a) Dimensionless mean ($H_\infty/W$) and (b) minimum ($H_{min}/W$) film thickness, as a function of Ca. The black curve represents the best fit curve obtained using the \textit{Taylor's law} model with $P=0.544,Q=2.061$ for  $H_\infty/W$ and with $P=0.372, Q=1.247$ for $H_{min}/W$. For the mean film thickness, predictions based on the coefficients $P,Q$ from the \textit{Taylor's law} (\citep{aussillous2000quick}, \citep{klaseboer2014extended} and \citet{zhu2016pancake}) are also shown.}\label{fig:exp3}
\end{figure}

Motivated by the qualitative agreement for the mean film thickness value between the experimental data and the 3D BIM simulations for pressure-driven droplets (figure \ref{fig:exp3}(\textit{a})), we perform a 3D BIM simulation using the solver developed in \citet{zhu2016pancake}, suitably adapted for gravity-driven droplets. Details of the numerical scheme are referred to that paper. 
\subsection{Comparison with 3D simulations}\label{NumericsComp}
The current numerical simulations only address the cases where the inner and outer viscosities are the same, namely $\lambda=1$. To realize it experimentally, three different drop volumes, 0.44ml, 1ml and 1.5ml, were injected in the Hele-Shaw cell resulting in $Ca$=0.032, 0.043 and $0.046$ with the corresponding $Bo$=1.81, 4.04 and 6.2. The error in volume injected decreased from $10\%$ to $3\%$ as we moved from the smallest to the largest drop volume. 

The experimental film thickness maps for the chosen $Ca$ range show that the precise shape of the pancake in-plane shape is close to an oval. Hence, the experimental in-plane drop shape is obtained by fitting the following equation 
\begin{equation}
	\frac{x^2}{(L/2)^2}+\frac{y^2 }{(T/2)^2} e^{cx}=0,
	\label{exp:shapePancake}
\end{equation}
on an instantaneous image of the drop where $c$ is a fitting parameter (figure \ref{fig:expMap}(\textit{a})). 
\begin{figure}
	\centering
	\begin{subfigure}{0.48\textwidth} 
		\includegraphics[trim=10 0 5 0,clip,width=\textwidth]{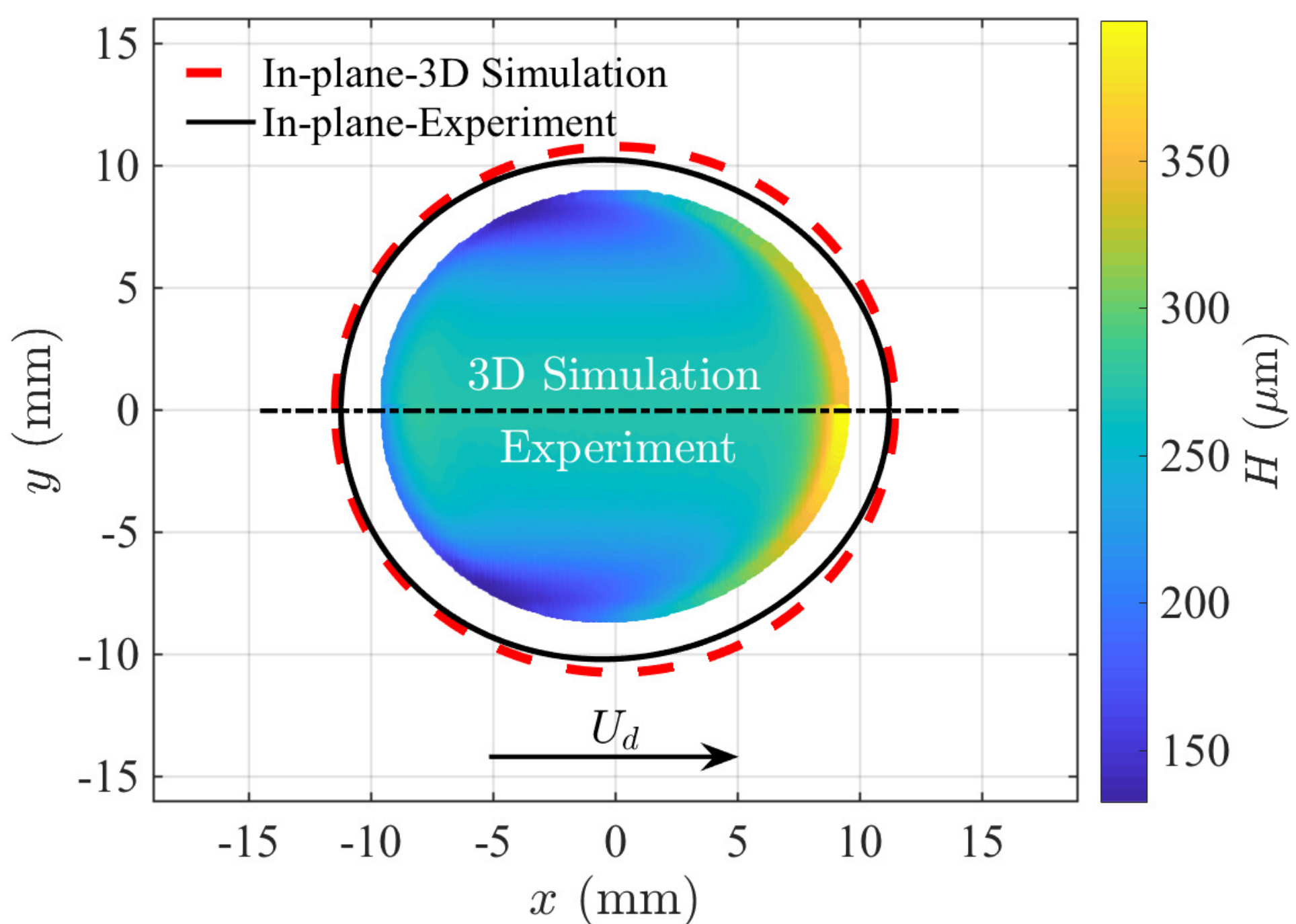}
		\caption{} \label{fig:exp4a}
	\end{subfigure}
	\begin{subfigure}{0.48\textwidth} 
		\includegraphics[trim=0 0 5 0,clip,width=\textwidth]{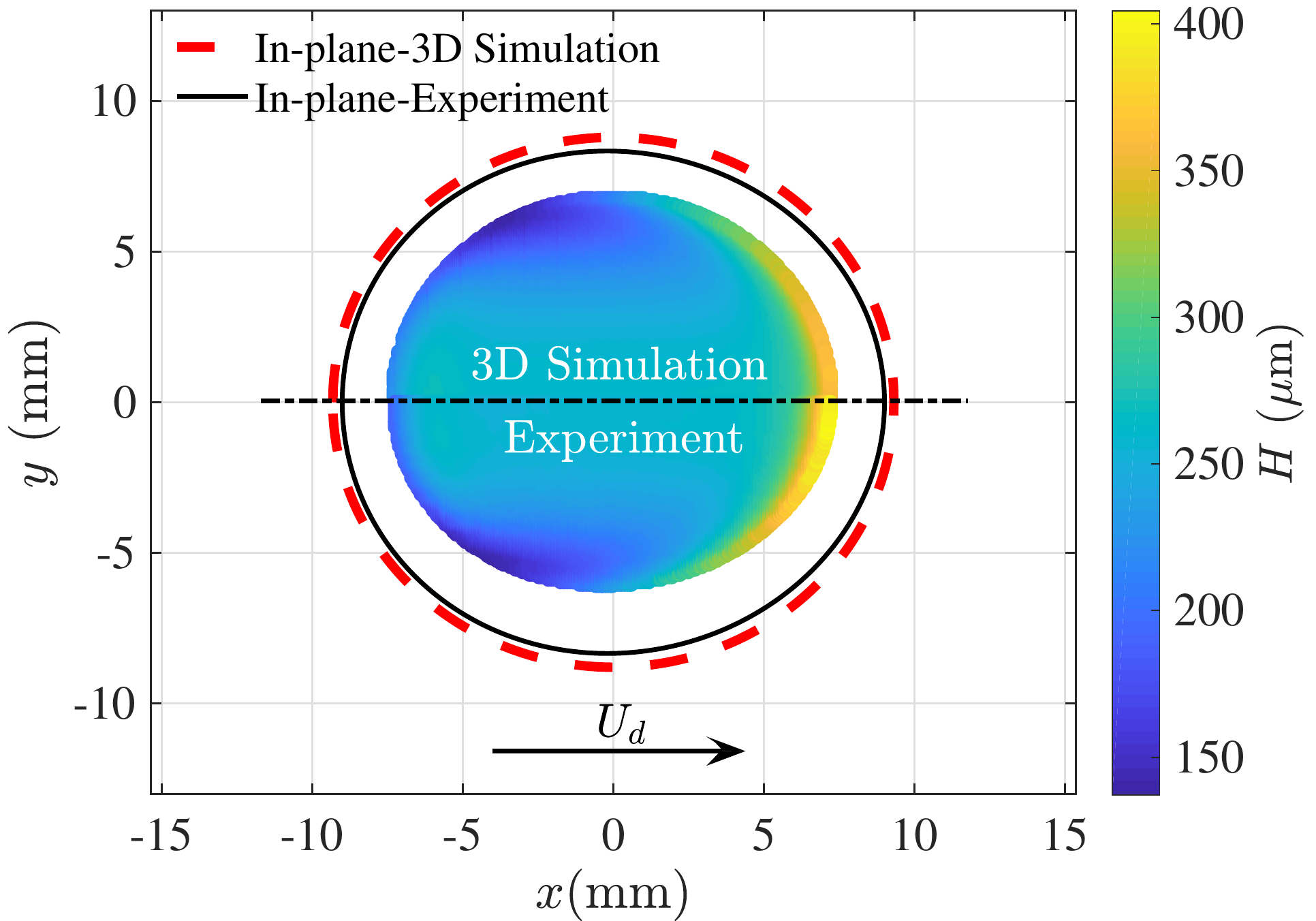}
		\caption{} \label{fig:exp4b} 
	\end{subfigure}
		\begin{subfigure}{0.48\textwidth} 
		\includegraphics[trim=0 0 5 0,clip,width=\textwidth]{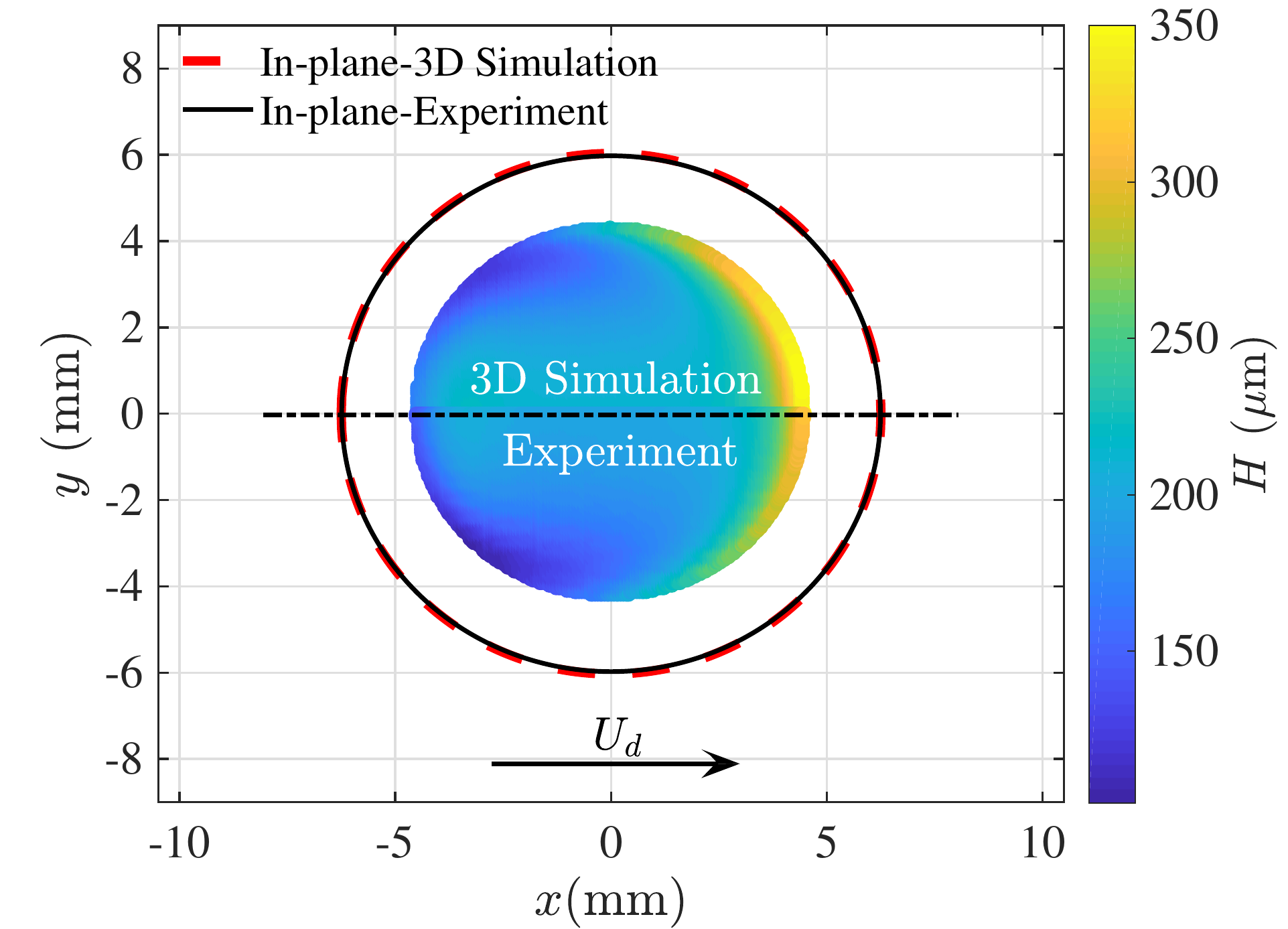}
		\caption{} \label{fig:exp4c} 
	\end{subfigure}
	\caption{Film thickness map whose top (resp. bottom) half corresponds to the 3D BIM (resp. experimental) data for three drop volumes (a) 1.5\ ml ($Ca=0.046,Bo=6.2$), (b) 1\ ml ($Ca= 0.043,Bo=4.04)$ and (c) 0.44\ ml ($Ca= 0.032,Bo=1.81$).The viscosity ratio $\lambda \approx 1$. The experimental and numerical in-plane shapes are represented by black and red dashed curves, respectively. The numerical results for $H_\infty$ along the centreline deviate with the experimental data by a factor of $2\%, 3\%$ and $5\%$ for the three cases, respectively.} \label{fig:exp4}
\end{figure}

Experimentally, due to large thickness gradient along the drop edges, the CCI sensor fails to capture the thickness in these regions. Thus the map is obtained for an area smaller than the in-plane shape of the drop (black curve in figure \ref{fig:exp4}). On the contrary, the numerical simulations are capable of retrieving the complete film thickness map, but for making a visually effective comparison between the experiments and numerics, only the part of the numerical result, with the same area as the experimental data, is shown in figure \ref{fig:exp4}. It's top/bottom half corresponds to the numerical/experimental data. The red dashed curve refers to the numerical in-plane shape of the drop. 

Both the experiments and simulations capture the formation of catamarans at the lateral transition regions, a uniform film thickness in the centre and a very high film thickness at the front edge of the drop. The agreement is almost quantitative. The relative error in the uniform film thickness $H_\infty$ for drop volumes 0.44\ ml, 1\ ml and 1.5\ ml, is  $5\%, 3\%$ and $2\%$ with absolute values as $12\ \mu$m, $7\ \mu$m and $6\ \mu$m, respectively. 

The numerical solution is further validated by making several streamwise (figure \ref{fig:exp6}) and spanwise (figure \ref{fig:exp5}) cuts along the largest drop of volume $1.5$\ ml. Along the centreline, the 3D BIM simulation captures precisely the lubrication film variation: large film thickness at front edge, followed by a constant thickness profile, ending in a small oscillation before posing an increasing trend at the rear edge. There is a good quantitative comparison between the experiments and numerics, with a slight variation in the film thickness along the advancing meniscus. 
\begin{figure}
\centering
\includegraphics[width=0.85\linewidth]{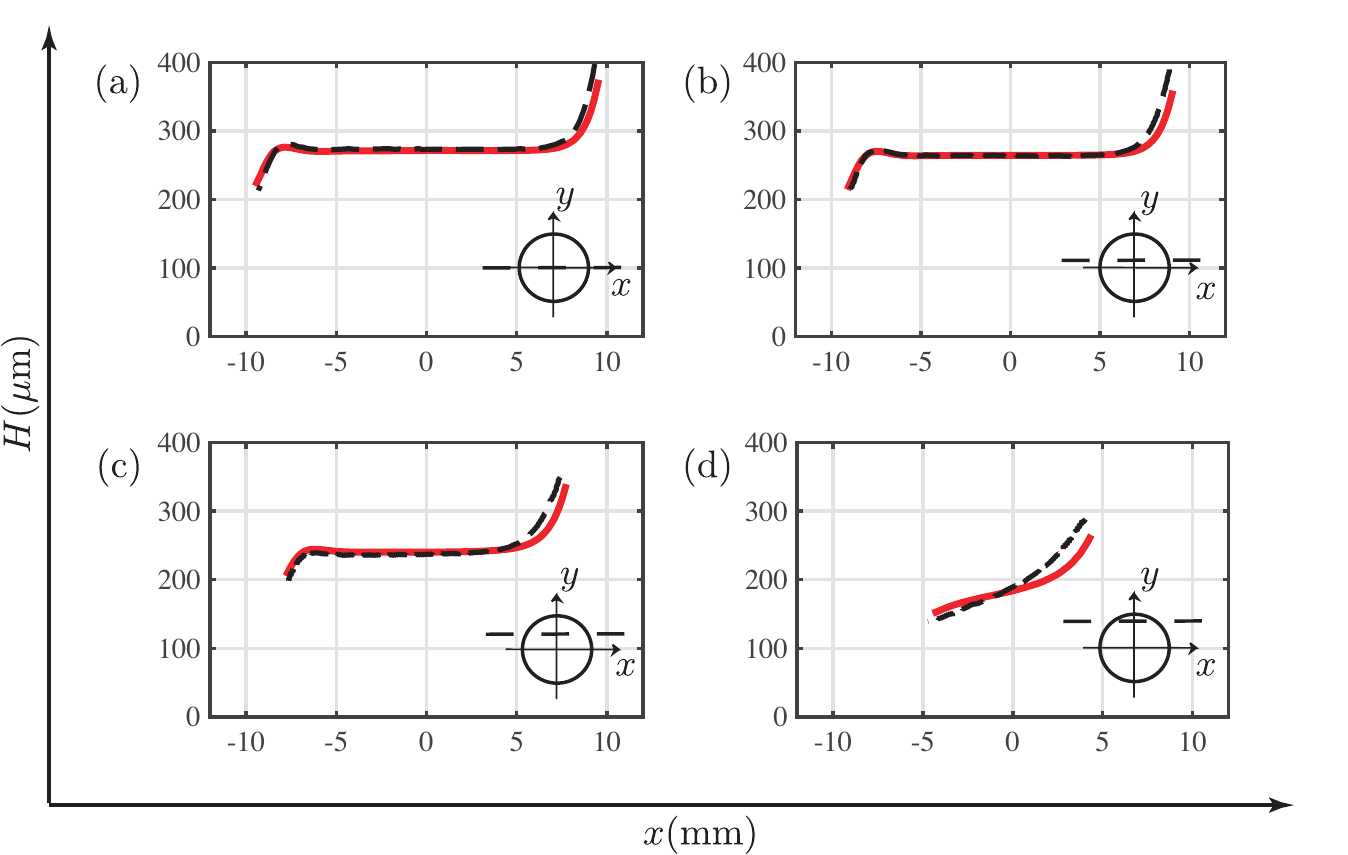}
\caption{Film thickness cuts made along the streamwise directions at (a) $y=0$, (b) $y=2.5$ mm, (c) $y=5$ mm and (d) $y=7.5$ mm, where $Ca=0.046,Bo=6.2$. Black dashed lines represent the experimental results and red lines the numerical predictions. The decrease in the film thickness towards the lateral edges can be observed by comparing (a) and (d) where the mean film  thickness decreases by around $30\%$ signifying the appearance of catamarans 
close to $(x,y)\approx (-4,7.5)$\ mm.}
\label{fig:exp6}
\end{figure}
\begin{figure}
\centering
\includegraphics[width=1\linewidth]{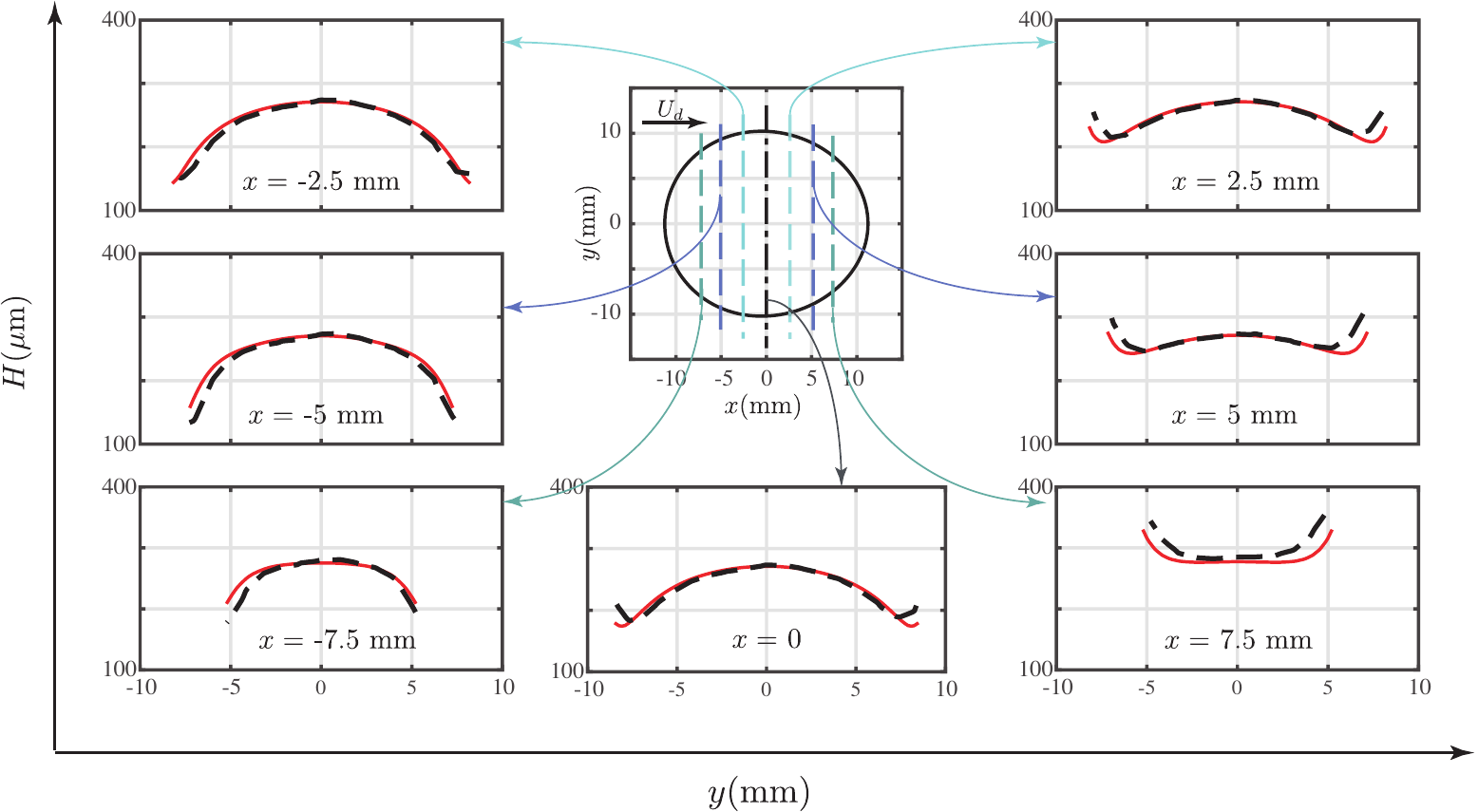}
\caption{Film thickness variation along the spanwise 
directions at $x= 0$, $x= \pm 2.5$ mm, $x= \pm 5$ mm and $x= \pm 7.5$ mm, where 
$Ca=0.046,Bo=6.2$. Black dashed lines represent the experimental results and red 
lines the numerical results. Transverse cuts enclosed by the region 
$x=-2.5$ mm to $x=-5$ mm highlight the minima in the lubrication film along the lateral 
edges.}
\label{fig:exp5}
\end{figure}%
\section{Analysis of the film thickness pattern \label{pancakes_tx:2D}}

In order to rationalise the film thickness pattern observed in \S \ref{NumericsComp}, we model hereafter the problem using a lubrication approach. For simplicity, we formulate the 2D lubrication equation assuming the drop dynamic viscosity $\mu_i = 0$. 
\subsection{Formulating the nonlinear 2D lubrication equation} \label{2dnonlinear}
Applying the long-wavelength assumption \citep{oron1997long} and by neglecting inertia, the 2D nonlinear lubrication equation (see details in Appendix \ref{2d_derivation}) for the film thickness ${H}$ separating the interface from the wall, in the reference frame moving at the drop velocity $U_d$, can be derived. Using the pancake radius $a$ as the characteristic length and $a/U_d$ as the characteristic time, the dimensionless lubrication equation for the steady profile in the dimensionless coordinate system $\bar{x}, \bar{y}$ is written as
\begin{equation}
	\frac{\partial}{\partial \bar x} \left[ \bar{H}^3 \left( \frac{1}{3 Ca} \bar{\kappa}_{x} - \frac{Bo}{3 Ca} \right) - \bar{H}\right]+ \frac{\partial}{\partial \bar y} \left( \bar{H}^3 \frac{1}{3 Ca} \bar{\kappa}_{y} \right)=0,
	\label{pancakes_eq:lub2Dsteady}
\end{equation}
where ${\bar{\kappa}}$ is the mean curvature of the interface, given by $\bar{\kappa}=\nabla \cdot {\mathbf{n}}$, where the unit normal vector ${\mathbf{n}}$ on the interface is given by 
\begin{equation}
	{\mathbf{n}}=\frac{\left( -\bar{H}_x,-\bar{H}_y,1\right)^T}{\sqrt{1+\bar{H}_x^2+\bar{H}_y^2}}.
\end{equation}

Note the anisotropy of the fluxes in equation \eqref{pancakes_eq:lub2Dsteady}: both the buoyancy and the motion in the 
$\bar{x}$ direction do not affect the flux in the $\bar{y}$ direction, breaking the isotropy induced by the capillary pressure gradient. 

The nonlinear equation \eqref{pancakes_eq:lub2Dsteady} together with the equation for the interface curvature $\bar{\kappa}$ are solved numerically by the commercial finite element solver COMSOL Multiphysics. The two variables for this coupled system of partial differential equations are $\bar{H}$ and $\bar{\kappa}$. As boundary conditions we impose the film thickness $\bar{H} = W/2a$ and the mean curvature $\bar\kappa= \bar\kappa_{f,r}$ at the droplet mid height. The mean curvature boundary condition in the static meniscus is composed by a component in the $(\bar r,\theta)$-plane and a component in the $(\bar r,\bar z)$-plane. In the spirit of \citet{meiburg1989bubbles} and \citet{Nagel:200506}, we consider the local capillary number defined with the normal velocity to the static cap for the mean curvature boundary condition model:
\begin{equation}
	\bar \kappa_{f,r}(\bar r,\theta) = \underbrace{\frac{2a}{W}\Bigg( \frac{1 + T_{f,r} (3 Ca |\cos \theta |)^{2/3}}{1+Z_{f,r} (3 Ca |\cos \theta |)^{2/3}}}_{\text{$(\bar r,\bar z)$-plane}}\Bigg) + \underbrace{\frac{\pi}{4} \frac{1}{\bar r}}_{\text{$(\bar r,\theta)$-plane}},
	\label{pancakes_eq:kappaCorrection2D}
\end{equation}
where the coefficients with subscript $f$ have to be used for $\theta \in [-\pi/2,\pi/2]$ and the ones with subscript $r$ for $\theta \in [\pi/2,3\pi/2]$, where $\theta$ and $\bar r$ are defined as $\theta=\arctan(\bar{y}/\bar{x})$ and $\bar r=(\bar{x}^2+\bar{y}^2)^{1/2}$, respectively. The values of the coefficients are $T_f=2.285$, $T_r=-0.5067$, $Z_f=0.4075$ and $Z_r=-0.1062$. The curvature boundary condition model in the $(\bar r,\bar z)$-plane is inspired by the equivalent model of \citet{balestra2018viscous}, which has been developed by an extensive study for the 2D planar Stokes problem. The validity of this model has recently been corroborated by \citet{atasi2018measure} for pancake bubbles. The correction $\pi/4$ for the in-plane curvature $1/\bar r$ in the $(\bar r,\theta)$-plane, where $\bar r=1$ for a circular geometry, has been derived asymptotically by \citet{park1984two}. Note that a more involved model could be used to describe the out-of-plane curvature (($\bar{r},\bar{z}$)-plane) in the lateral transition regions \citep{burgess1990analysis}. 
\begin{figure}
\centerline{\includegraphics[width=0.9\textwidth]{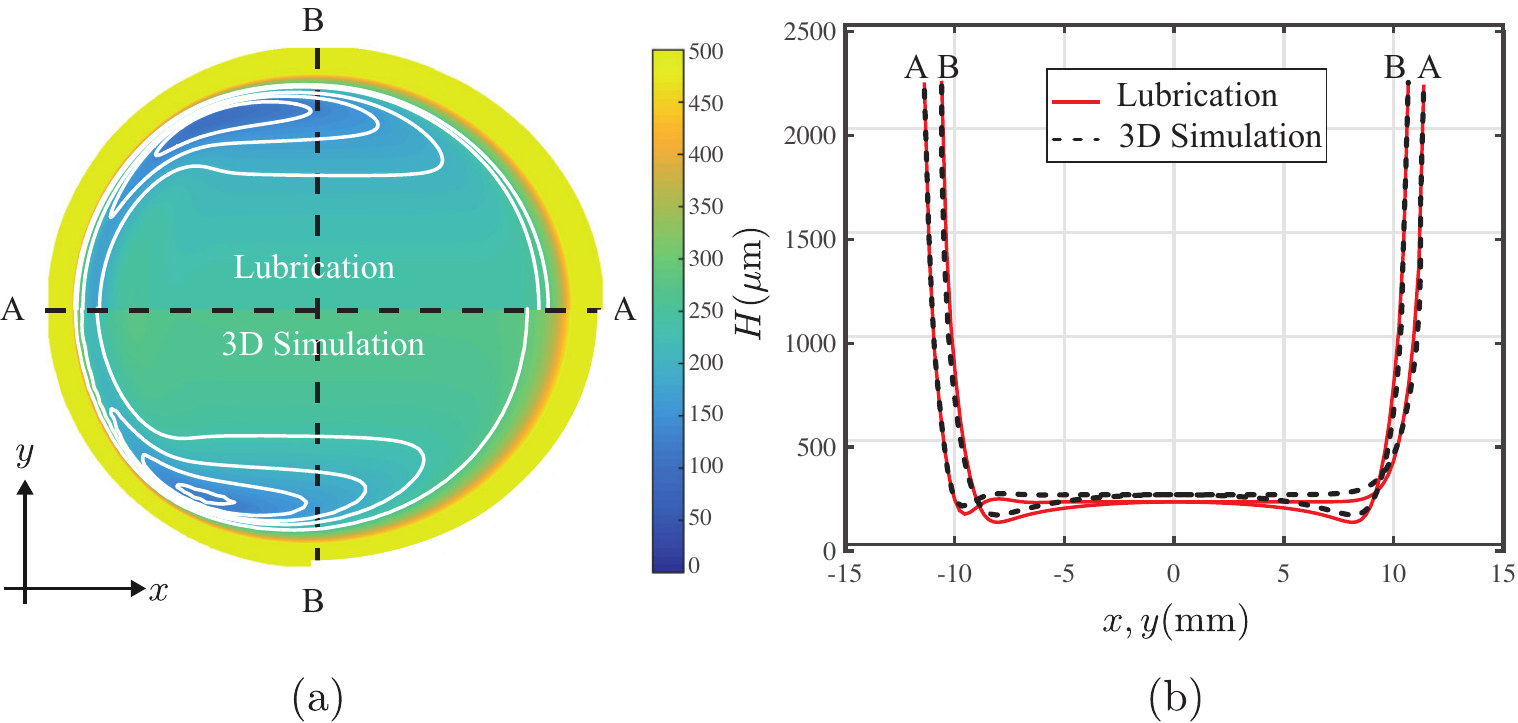}}
\caption{(a) Comparison between the solution of the nonlinear lubrication equation assuming $\lambda=0$ (top half) and that of the 3D BIM simulations assuming $\lambda=1$ (bottom half), where $Ca = 4.6 \times 10^{-2}$ and $Bo = 6.2$. (b) Comparison for cuts made along the streamwise direction A-A and spanwise direction B-B.\label{pancakes_fig:compLubLailai} }
\end{figure}

In the present work we extract the pancake shape from the results of the 3D BIM simulations for $\lambda=1$. As explained in \S \ref{NumericsComp}, the in-plane boundaries of the deformed pancake in the $(\bar r,\theta)$-plane can be well described by equation \eqref{exp:shapePancake}. 

It has to be stressed that the used lubrication equation should not, a priori, be used in the static meniscus region close to the boundary, where the interface slope is large. However, we have found that such an approach gives surprisingly good results if one uses the model for the static rim curvature \eqref{pancakes_eq:kappaCorrection2D} for the curvature boundary condition (see \citet{BalestraPhD} for a discussion of the axisymmetric case), which directly sets the film thickness profile in that region. Hence, such an approach can be used to numerically obtain the film thickness profile over the entire domain, also behind its validity range.

The comparison between the film thickness profile obtained by the solution of the nonlinear lubrication equation using the model equation \eqref{pancakes_eq:kappaCorrection2D} for the static cap mean curvature $\bar \kappa_{f,r}(r,\theta)$, with the one obtained by the 3D BIM simulations, is shown in figure \ref{pancakes_fig:compLubLailai}. One can observe that both methods predict the formation of catamarans at the lateral transition regions, a uniform film thickness in the centre and oscillations at the back. 
In spite of the strong assumptions made for this model, the agreement is surprisingly good, even with an iso-viscous drop ($\mu_i=\mu_o$).
The relative error in the uniform film thickness is of $10\%$ and its absolute value is $30\, \mu$m. The thin-film pattern shown by both approaches, as well as by the experiments, is therefore indeed a robust feature. Supported by this agreement, we investigate the thin-film pattern using the linearized version of this simple 2D lubrication model, which is computationally much cheaper than the 3D Stokes simulations. 
\subsection{Qualitative analysis of thickness pattern using the linearized 2D lubrication equation \label{pancakes_tx:2Dlin}}
The qualitative nature of the film thickness pattern can be inferred by performing a linear analysis of the 2D lubrication equation \eqref{pancakes_eq:lub2Dsteady}.
With the use of the film thickness decomposition $\bar{H} = \bar H_{\infty} + \varepsilon h$, where $\bar H_\infty = H_\infty/ a$, the linear equation for the first-order film thickness correction reads:

\begin{equation}
	\frac{\bar H_{\infty}^3}{3Ca} (\underbrace{h_{\bar{x}\bar{x}\bar{x}\bar{x}}+2 h_{\bar{x}\bar{x}\bar{y}\bar{y}}+h_{\bar{y}\bar{y}\bar{y}\bar{y}}}_{\Delta^2 h}) - \left( 1 +\frac{\bar H_{\infty}^2 Bo}{Ca}\right) h_{\bar{x}} = 0.
	\label{pancakes_eq:lub2Dlin}
\end{equation}
The film thickness $\bar{H}_{\infty}$ is expressed using the empirical model \eqref{HAsympExp} (\cite{taylor1961deposition}, \cite{aussillous2000quick} and \cite{balestra2018viscous}), with $P = 0.643$ and $Q = 2.2$.

Equation \eqref{pancakes_eq:lub2Dlin} for the film thickness correction around the uniform film thickness $\bar H_{\infty}$ can be solved as a boundary-value problem, as recently conducted by \citet{atasi2018measure}. In contrast to the nonlinear solution of Sec.\ \ref{2dnonlinear}, here we only solve the lubrication equation from the thin-film region up to the beginning of the dynamic meniscus region. This is equivalent to looking at the first-order correction of the uniform thin film region due to the matching of the film thickness in the dynamic meniscus region to a larger value. In the present context, we impose a film thickness correction $h=A$ and a mean curvature of the order $\Delta h = 1/A + 1/\bar r$ on the perimeter, with $A$ as a constant value of $10^{-3}\times\bar{H}_{\infty}$. This boundary condition does not have to be understood as a rigorous matching approach, but rather as a way to find the structure of the film thickness profile in the region where it is close to be uniform. A rigorous matching for the limit $Ca\ll1$, can be found in \citet{park1984two}. The maps of the film thickness correction $h$, together with some profiles along the streamwise and spanwise directions, are shown in figure \ref{pancakes_fig:2Dlinear}. 
\begin{figure}
\centerline{\includegraphics[width=0.85\textwidth]{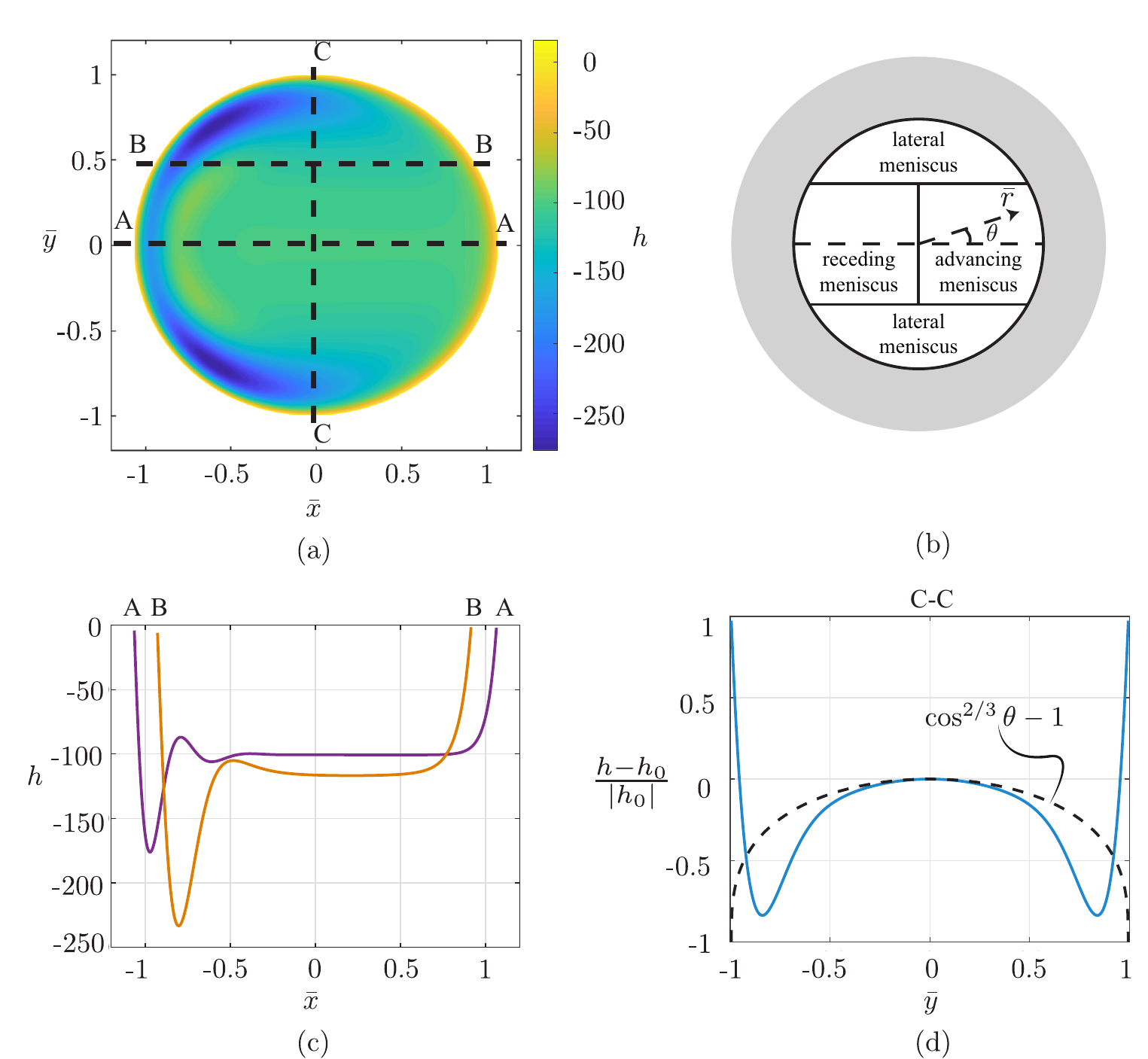}}
\caption{Linear film thickness 
correction $h$ around the uniform film thickness $\bar{H}_{\infty}$ for a 
pancake droplet (a,c,d). The film thickness correction map is shown in (a), 
the cuts along the streamwise direction at two different $\bar y$-locations are 
plotted in (c) and the normalized difference of the film thickness correction 
along the spanwise cut C-C with respect to $h_0 = h(\bar y=0)$ along this cut is 
shown in (d), where the law $(\cos \theta)^{2/3}$ is indicated by the black dashed line. $A = 10^{-3}\times\bar{H}_{\infty}$, $Ca = 4.6 \times10^{-2}$, $Bo 
=6.2$ and $\alpha=2.2$. The polar coordinates $(\bar r,\theta)$ are introduced and the boundaries are 
highlighted by the grey area in (b).}\label{pancakes_fig:2Dlinear}
\end{figure}

First, it can be clearly observed that the linear lubrication equation with a perturbed film thickness and curvature along the domain boundary is able to reproduce the catamaran-like pattern observed in pancake droplets as seen in \S \ref{NumericsComp} and \S \ref{2dnonlinear}. The film thickness is the smallest in the lateral part of the pancake (see figure \ref{pancakes_fig:2Dlinear}(\textit{a})), so that its 3D shape resembles the hull of a catamaran. Therefore, we can conclude that this pattern is the generalization of the one-dimensional oscillations found by \cite{bretherton1961motion} at the rear of an axisymmetric bubble for a 2D concave structure, like a pancake droplet, and is intrinsically related to the anisotropy of the equation. 

Second, the film thickness correction along the streamwise direction $\bar{x}$ (see figure \ref{pancakes_fig:2Dlinear}(\textit{c})) deviates from a uniform profile as expected from Bretherton's theory \citep{bretherton1961motion}. The film thickness oscillates at the rear meniscus and increases monotonically at the front meniscus. Note that the film thickness correction in the uniform film region of a pancake is not vanishing as the base film thickness $\bar{H}_{\infty}$ is given by equation \eqref{HAsympExp}, which is an asymptotic estimate for $Bo=0$ but not an exact solution of the lubrication equation with $Bo \neq 0$. Furthermore, it can be observed that the more one moves away from $\bar y=0$, the more the thickness of the film is reduced. Therefore, the thickness of the film left by the front meniscus is not uniform.

To better highlight this crucial point, we show in figure \ref{pancakes_fig:2Dlinear}(\textit{d}) the normalized difference between the film thickness correction and its value at $\bar{y}=0$. The film thickness decreases as $|\bar{y}|$ increases, before increasing again close to the edge to match the boundary condition. 

These qualitative observations can be rationalised by simplifying the linear lubrication equation \eqref{pancakes_eq:lub2Dlin} for the different regions of the domain (see figure \ref{pancakes_fig:2Dlinear}(\textit{b})). The lubrication equation \eqref{pancakes_eq:lub2Dlin} in polar coordinates can be simplified to
\begin{equation}
	\frac{\bar H_{\infty}^3}{3Ca} h_{\bar r \bar r \bar r \bar r} - \left( 1 +\frac{\bar H_{\infty}^2 Bo}{Ca}\right) \left( \cos \theta  h_{\bar r} - \frac{\sin \theta}{\bar r} h_{\theta}\right) = 0.
	\label{pancakes_eq:lub2DlinPolar}
\end{equation}

For small polar angles $\theta$, the contribution $ (\sin \theta/\bar r)  h_{\theta} $, which corresponds to the flux in the tangential direction, can be neglected so that the linear lubrication equation becomes, after integration along $\bar r$:
\begin{equation}
	 h_{\bar r \bar r \bar r} = K_p  h,
	\label{pancakes_eq:lub2DlinPolarFrontRear}
\end{equation}
with
\begin{equation}
	K_p = \left( \frac{3Ca} {\bar H_{\infty}^3} +\frac{3 Bo}{\bar H_{\infty}}\right) \cos \theta,
	\label{pancakes_eq:Kp}
\end{equation}
which is the linearised one dimensional equation for the Landau-Levich-Derjaguin-Bretherton problem (\cite{landau1988dragging}, \cite{derjaguin1943thickness} and \cite{bretherton1961motion}) in the radial direction $\bar r$ projected on the streamwise direction. Therefore, we know from the solution of  \citet{bretherton1961motion} that the film thickness is oscillating at the rear meniscus and monotonically increasing at the front one. Focusing for now on $Bo = 0$, we know that the thickness deposited by a front meniscus depends on the velocity normal to the interface. In this case, one has therefore $\bar H_{\infty} \sim Ca_p^{2/3}$ with $Ca_p = Ca \cos \theta$ as the local capillary number at a given polar angle $\theta$. Hence, the film thickness in the central region of the pancake varies like $(Ca \cos \theta)^{2/3}$. Similar results have been reported for a pressure-driven red blood cell traversing in a non-axisymmetric passage \citep{halpern1992squeezing} and in pancake droplets by \citet{reichert2018topography}. Once a given film thickness is set by the front meniscus, the same thickness will be present over the entire thin film region at the corresponding spanwise location $\bar y$. The good agreement between the dependence in $({\cos\theta})^{2/3} $ of the film thickness and the profile along the spanwise direction obtained by resolving the 2D lubrication equation is shown in figure \ref{pancakes_fig:2Dlinear}(\textit{d}).

Similarly, the oscillations at the rear meniscus depend on the polar angle. For a pancake droplet, due to the film thickness nonuniformity resulting from the nonuniform deposition at the front, the wavelength of the oscillations at the back scales as $\lambda_{osc} \sim (Ca \cos\theta)^{1/3}$. Given the $1/3$ power-law dependence, the wavelength is almost unchanged, before rapidly reducing to $0$ when $\theta \rightarrow \pm \pi/2$ (see figure \ref{pancakes_fig:2Dlinear}(\textit{a})). 

It is important to note that a plane cut of the film thickness at a given angle $\theta$ does not present a region of constant film thickness. A pancake droplet cannot be seen just as the collection of different one-dimensional profiles obtained by the solution to equation \eqref{pancakes_eq:lub2DlinPolarFrontRear} for different polar angles $\theta$. In fact, the film thickness at any spanwise location $\bar y$ is set by the front meniscus at the corresponding polar angle $\theta$ and equation \eqref{pancakes_eq:lub2DlinPolarFrontRear} only indicates the scaling of this film thickness as well as the oscillations at the back.

For $\theta \rightarrow \pm \pi/2$, which corresponds to the lateral meniscus region (see figure \ref{pancakes_fig:2Dlinear}(\textit{b})), the tangential flux term $ (\sin \theta/\bar r)  h_{\theta} $ in equation \eqref{pancakes_eq:lub2DlinPolar} can no longer be neglected. \citet{burgess1990analysis} performed an involved analysis of the lubrication equation in this region for a pancake droplet at low capillary numbers and found that the local film thickness in the so-called lateral transition regions scales as $Ca^{4/5}$ rather than as $Ca^{2/3}$. Therefore, for $Ca\ll 1$, the film thickness in these lateral regions is much smaller than the one in the other regions. This explains why one observes catamaran-like structures in the lateral regions of pancake droplets. Note that for the $Ca$-range considered in the present study, the film thickness in the lateral transition regions is still sufficiently large so that the viscous dissipation can be neglected also in these regions, as confirmed by the results of Sec.\ \ref{ExpNumRes}. Furthermore, \citet{burgess1990analysis} have also shown that the polar extent of these lateral regions scales as $Ca^{1/5}$, whereas their radial extent scales as $Ca^{2/5}$ instead of as $Ca^{1/3}$ that one has for the length of rear and front dynamic menisci of axisymmetric droplets (see also \citet{hodges2004sliding}).  
\section{Conclusions and perspectives \label{conclusion}}
We report the velocity, mean film thickness and thickness map for a droplet moving due to buoyancy in a vertical Hele-Shaw cell. The mean drop velocity compares well with the leading order velocity expression of \citet{gallaire2014marangoni}. This signifies that buoyancy and viscous drag force are the dominant forces in our experimental parameter range with the viscous dissipation in the film thickness and in the dynamical meniscus having a negligible effect on the droplet velocity. On the contrary, the dimensionless mean film thickness data was dependent on the dimensionless droplet velocity, expressed as $Ca$, and was fitted with the \textit{Taylor's law} model \citep{aussillous2000quick}. 

We also obtained the complete film thickness maps using a CCI optical pen mounted on a linear translation stage. Based on a boundary integral method, 3D Stokes equations were solved. These numerical results for $\lambda=1$ were in very good agreement with our experimental results. The thickness pattern had a distinct catamaran-like shape as experimentally observed for pressure-driven flows in \citet{huerre2015droplets} and \citet{reichert2018topography}. 

To understand the nature of the thickness patterns observed experimentally and numerically, the problem was approached using a lubrication equation, which was solved as a boundary value problem, rather than as an initial value problem, as recently conducted by \citet{atasi2018measure}. In spite of all the crude assumptions done for developing the model, its nonlinear solution for the film thickness profile of a pancake bubble compared surprisingly well with the results of 3D BIM simulations, evidencing the robustness of the thin-film pattern.

In order to unravel the structure of the film thickness profile, we linearized the lubrication model and solved for the linear thickness corrections around a uniformly-thick film. We have been able to show that not only the oscillations at the rear meniscus, but also the catamaran-like pattern can be directly retrieved by solving the linear 2D lubrication equation when perturbing the film thickness at the boundaries, which mimics the presence of a meniscus of greater film thickness. In particular, the catamaran-like structure results from the anisotropic flux induced by the motion of the walls with respect to the pancake and the need to match the film thickness to larger values in the dynamic meniscus region surrounding the region where the thin film is rather uniform. This pattern is therefore independent of the force driving the motion. In fact, in totally different contexts, the same pattern is also found in drops levitating on a moving substrate (\citealp{hodges2004sliding,lhuissier2013levitation}) as well as in oleoplaning drops (\citealp{daniel2017oleoplaning}). In the central part of the pancake droplet, the thickness left by the front meniscus scales as $(Ca \cos \theta)^{2/3}$, and depends therefore on the velocity normal to the interface. This scaling no longer holds in the lateral transition region, where the component of the flux tangential to the interface becomes important and the thickness of the film is much smaller, resulting in the formation of the catamaran-like structure.

Finally, we would like to highlight a contrasting feature seen between drops moving in a cylindrical tube and that in a Hele-Shaw cell. In cylindrical tubes the main difference between the pressure-driven and buoyancy-driven bubble motion, as found by \citet{bretherton1961motion}, is that buoyant bubbles may remain stuck if the capillary radius is less than $0.918l_c$, where the capillary length $l_c=\sqrt{\gamma/\Delta\rho g}$. This failure results from the impossibility to match the static gravity-corrected meniscus shape with the flat thin film region. A similar result was obtained recently in the planar geometry by \citet{lamstaes2017arrested} with a prefactor of $0.847$. Interestingly enough, we suspect that there is no such bubble arrest in Hele-Shaw cells, as a consequence of the additional direction which adds a degree of freedom in the curvature. As shown in figure \ref{g4}, our experimental results show marked drop motion when the half cell gap is below $0.847l_c$.
\\ 
\\
\\
\begin{figure}
\centering
\includegraphics[width=0.65\textwidth]{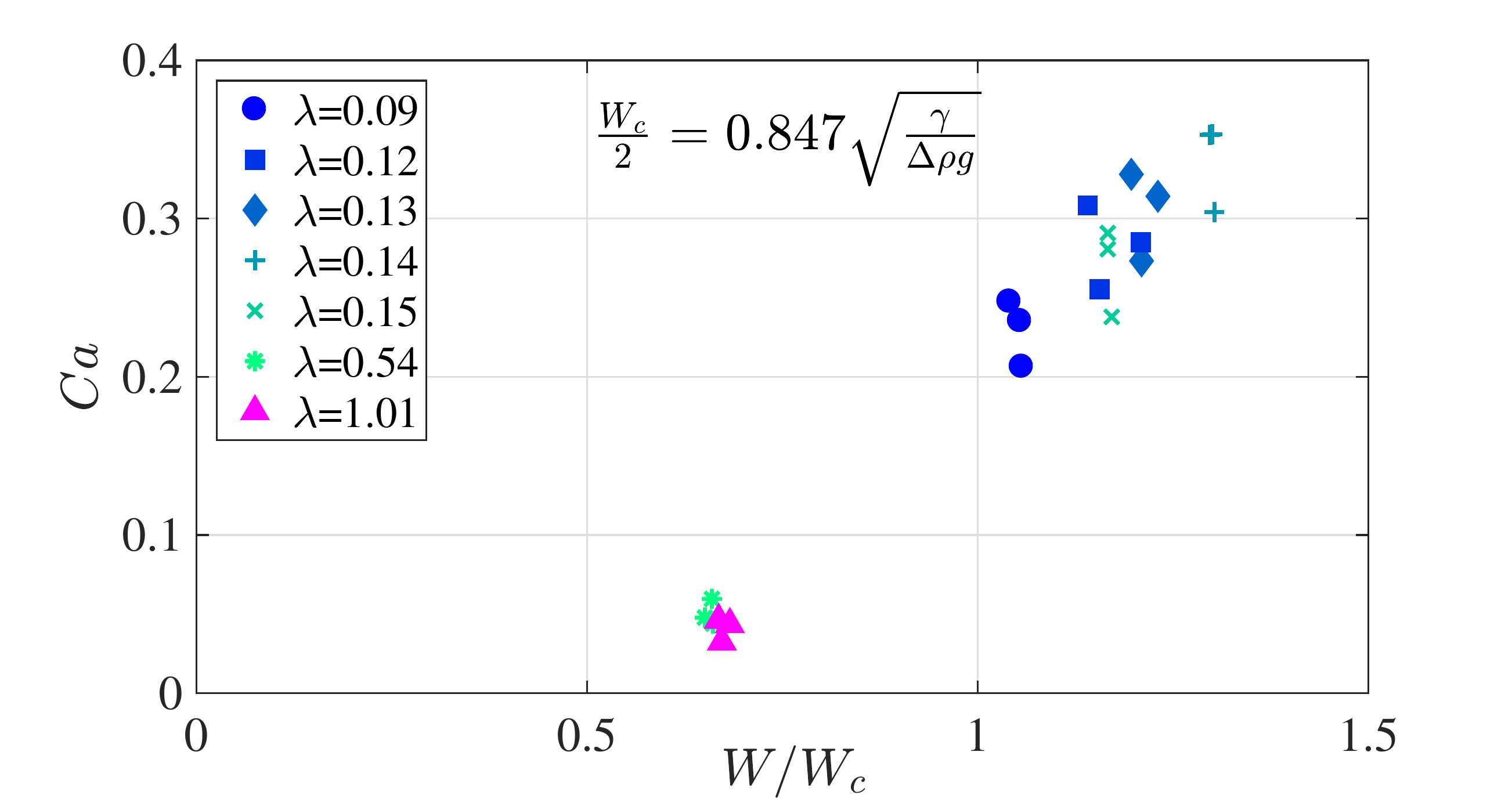}
\caption{ As stated by \citet{lamstaes2017arrested}, for a planar geometry the drops should get stuck when $W<W_c$. Our experiments show that the pancake-shaped drops continue to move beyond this limitation.}\label{g4}
\end{figure}
I. S. thanks the Swiss National Science Foundation (grant no. 200021-159957). The computer time is provided by the Swiss National Supercomputing Centre (CSCS) under project ID s603. An ERC starting grant ``SimCoMiCs 280117''  is gratefully acknowledged. L. Z. thanks a VR International Postdoc Grant ``2015- 06334'' from the Swedish Research Council. We thank Ludovic Keiser for fruitful discussions.
\appendix \label{appA}
\section{CCI working principle} \label{CCI}
We hereby describe the principle of the Confocal Chromatic Imaging technique. An achromatic lens decomposes the incident white light into a continuum of monochromatic images which constitutes the measurement range. The light reflected by a sample surface put inside this range is collected by a beam splitter. A pinhole allows then to block the defocused light that does not come from the sample surface. Eventually, the spectral repartition of the collected light is analyzed by a spectrometer. The wavelength of maximum intensity is detected and the distance value is deduced from a calibration curve. Several reflecting interfaces can be detected at the same time, allowing thickness measurement of thin transparent layers.
\section{Derivation of two-dimensional nonlinear lubrication equation for pancake droplets} \label{2d_derivation}
The derivation of the model equation presented in \S \ref{pancakes_tx:2Dlin} is briefly outlined hereafter. Considering the same physical properties for the droplet and the outer medium as outlined in \S \ref{introduction}, and under the assumption of negligible inertia \citep{oron1997long} with $\rho_i \ll \rho_o$, and $\mu_i \ll \mu_o$ the three dimensional momentum equations reads
\begin{align}
        0 = &- \frac{\partial p}{\partial x} + \mu_o  \Big(\frac{\partial^2 u}{\partial x^2} + 
        \frac{\partial^2 u}{\partial y^2} + \frac{\partial^2 u}{\partial z^2} \Big) -\rho_o  g \label{Th_eq:xmom},\\
        0 = &- \frac{\partial p}{\partial y} + \mu_o   \Big(\frac{\partial^2 v}{\partial x^2} + 
        \frac{\partial^2 v}{\partial y^2} + \frac{\partial^2 v}{\partial z^2} \Big) \label{Th_eq:ymom},\\
        0 = &- \frac{\partial p}{\partial z} +\mu_o  \Big(\frac{\partial^2 w}{\partial x^2} + 
        \frac{\partial^2 w}{\partial y^2} + \frac{\partial^2 w}{\partial z^2} \Big)\label{Th_eq:zmom}. 
\end{align}
Using $L$ as the characteristic length scale in $x$ and $y$ direction and the film thickness $H$ as the characteristic length scale in $z$ direction, the film aspect ratio $\epsilon$ is defined as $\epsilon = H/L$. The long wavelength approximation is employed since $\epsilon \ll 1$. Mass conservation indicates that the characteristic velocity in $z$ direction ($W$) is much smaller than the other two components ($U$ in $x$ and $V$ in $y$ direction), $W \sim \epsilon U \ll U$ and $W \sim \epsilon V \ll V$. The Stokes equation simplifies as  
\begin{align}
        0 = &- \frac{\partial p}{\partial x} + \mu_o  \frac{\partial^2 u}{\partial z^2} -\rho_o  g \label{Th_eq:xmom}, \\
        0 = &- \frac{\partial p}{\partial y} + \mu_o  \frac{\partial^2 v}{\partial z^2} \label{Th_eq:ymom}, \\
        0 = &- \frac{\partial p}{\partial z} \label{Th_eq:zmom}. 
\end{align}

Integrating equation \eqref{Th_eq:zmom} in $z$ and applying dynamic boundary conditions yields $p=p_0 - \gamma \kappa$  where ${\kappa}$ is the mean curvature of the interface. Integrating equation \eqref{Th_eq:xmom} and \eqref{Th_eq:ymom} twice in $z$ and considering $u(z=0)=-U_d$ and the zero-slip boundary condition $v(z=0)=0$ as well as the zero-shear-stress interface  $\partial u (z=H)/\partial z=0$ and $\partial v (z=H)/\partial z=0$ yields the velocity components:
\begin{align}
        u = & \frac{ (\gamma \kappa_x - \Delta \rho g)}{\mu_o}  \Big(H-\frac{z}{2}\Big)z - U_d \label{Th_eq:xvel},\\
        v = &\frac{\gamma \kappa_y }{\mu_o}  \Big(H-\frac{z}{2}\Big)z \label{Th_eq:yvel}.
\end{align}
Since the inner medium density $\rho_i \ll \rho_o$, we replace $\rho_o$ by $\Delta \rho$ in equation \eqref{Th_eq:xvel} where $\Delta \rho= \rho_o -\rho_i$ represents the density difference between the inner and outer fluid. Integrating equation \eqref{Th_eq:xvel} and \eqref{Th_eq:yvel} in $z$ from $0$ to $H$ yields the flux in $x$ and $y$ direction as
\begin{align}
        Q^u = & \frac{ H^3}{3\mu_o} (\gamma \kappa_x - \Delta \rho g) - U_d H \label{Th_eq:xQ},\\
        Q^v= &\frac{ H^3}{3\mu_o} \gamma \kappa_y   \label{Th_eq:yQ}.
\end{align}
Finally integrating the continuity equation and applying the Leibniz Integral rule and the kinematic boundary condition at the interface yield the mass conservation equation expressed as
\begin{equation}
\frac{\partial H}{\partial t} + \frac{\partial Q^u}{\partial x}+ \frac{\partial Q^v}{\partial y}=0 \label{Th_eq:MassCon1}.
\end{equation}
Introducing equation \eqref{Th_eq:xQ} and \eqref{Th_eq:yQ} in equation \eqref{Th_eq:MassCon1} finally yields the lubrication  equation:
\begin{equation}
\frac{\partial H}{\partial t} + \frac{\partial}{\partial x}\Bigg[\frac{ H^3}{3\mu_o} (\underbrace{\gamma \kappa_x}_{\mathbf{I}} - \underbrace{\Delta \rho g}_{\mathbf{II}}) - \underbrace{U_d H}_{\mathbf{III}} \Bigg]
+ \frac{\partial}{\partial y} \Bigg[\frac{ H^3}{3\mu_o} \underbrace{\gamma \kappa_y}_{\mathbf{I}} \Bigg]=0. \label{Th_eq:MassCon2}
\end{equation}
The terms $\mathbf{I}$ in the spatial variation of the flux corresponds to the surface tension effects, term $\mathbf{II}$ to the variation due to the buoyancy force and term $\mathbf{III}$ accounts for the reference frame moving with the drop. Note the anisotropy of the fluxes: both the buoyancy and the motion in the $x$ direction do not affect the flux in the $y$ direction, breaking the isotropy induced by the capillary pressure gradient.

Using the pancake radius $a$ as the characteristic 
length and $a/U_d$ as the characteristic time, the dimensionless lubrication 
equation for the steady profile in the dimensionless coordinate system $\bar{x}, 
\bar{y}$ is written as equation \eqref{pancakes_eq:lub2Dsteady}.
\bibliographystyle{jfm}
\bibliography{bibliography}

\end{document}